\documentclass[
reprint,
superscriptaddress,
amsmath,amssymb,
prx,
floatfix,
longbibliography
]{revtex4-2}

\usepackage{graphicx}
\usepackage{bm}

\usepackage[T1]{fontenc} 
\usepackage{color,soul}
\usepackage[dvipsnames]{xcolor}

\usepackage{float}

\usepackage[skip=0pt plus0pt, indent=10pt]{parskip}

\usepackage{setspace}

\newcommand{\smallminus}{\scalebox{0.6}[0.6]{$-$}}
\newcommand{\smallplus}{\scalebox{0.6}[0.6]{$+$}}
\newcommand{\fwd}{^{\smallplus}}
\newcommand{\bck}{^{\smallminus}}

\begin{document}

\preprint{APS/123-QED}

\title{Activating high-power parametric oscillation in photonic-crystal resonators
}

\author{Grant M. Brodnik}
    \email[Correspondence email address: ]{grant.brodnik@nist.gov}
     \affiliation{Time and Frequency Division, National Institute of Standards and Technology, Boulder, CO, USA}
     \affiliation{Department of Physics, University of Colorado, Boulder, CO, USA}
\author{Lindell M. Williams}
     \affiliation{Time and Frequency Division, National Institute of Standards and Technology, Boulder, CO, USA}
     \affiliation{Department of Physics, University of Colorado, Boulder, CO, USA}
\author{Haixin Liu}
     \affiliation{Time and Frequency Division, National Institute of Standards and Technology, Boulder, CO, USA}
     \affiliation{Department of Physics, University of Colorado, Boulder, CO, USA}
\author{David R. Carlson}
     \affiliation{Time and Frequency Division, National Institute of Standards and Technology, Boulder, CO, USA}
     \affiliation{Octave Photonics, Louisville, CO}
\author{Atasi Dan}
     \affiliation{Time and Frequency Division, National Institute of Standards and Technology, Boulder, CO, USA}
     \affiliation{Department of Physics, University of Colorado, Boulder, CO, USA}
\author{Jennifer A. Black}
     \affiliation{Time and Frequency Division, National Institute of Standards and Technology, Boulder, CO, USA}
     \affiliation{Currently with Vescent Photonics, Golden, CO, USA}
\author{Scott B. Papp}
     \affiliation{Time and Frequency Division, National Institute of Standards and Technology, Boulder, CO, USA}
     \affiliation{Department of Physics, University of Colorado, Boulder, CO, USA}

\date{\today}

\begin{abstract} 
\noindent By engineering the mode spectrum of a Kerr microresonator, we selectively activate nonlinear phase matching amongst broadband parametric gain. At threshold, optical parametric oscillators (OPOs) emerge from vacuum fluctuations in the presence of a pump laser, and above threshold, OPOs seed the formation of intraresonator patterns and states, such as chaos and solitons. These competing nonlinear processes hinder an important application of OPOs as wavelength-variable, low-noise sources. Recently, nanopatterned microresonator OPOs have leveraged photonic crystal bandgaps to enable universal phase matching and control of nonlinear interactions. Here, we explore a design paradigm optimized for high-output power that uses geometric dispersion to suppress nonlinear interactions and a photonic crystal bandgap to activate only a single OPO interaction. Our devices convert an input pump laser to output signal and idler waves with powers exceeding 40 mW while maintaining spectral purity and side-mode suppression ratios greater than 40 dB.  We show that this approach suits custom wavelengths by measuring four independent oscillators that vary only photonic crystal parameters to select output waves. Our experiments demonstrate that microresonators functionalized by photonic crystals offer a versatile and lossless palette of controls for nonlinear laser conversion.

\end{abstract}

\maketitle

\noindent Nonlinear photonics continues to drive versatility in compact optical sources. For example, four-wave-mixing interactions are key in realizing entangled photons for quantum communications \cite{lu_chip-integrated_2019}, squeezed states for improvements in sensing \cite{Yang_2021}, and Kerr microresonator frequency combs \cite{delhaye_optical_2007,pasquazi_micro-combs_2018} and solitons \cite{herr_temporal_2014}. Recently, microresonator wavelength converters that leverage optical parametric oscillation (OPO) based on Kerr four-wave-mixing (FWM) have begun to address demand for coherent, constant-wave sources that span visible and near-IR wavebands \cite{lu_visible_2020,domeneguetti_parametric_2021,sun_advancing_2024}. Engineering the group-velocity dispersion (GVD, or dispersion) of microresonator OPOs through changes in geometry, e.g., waveguide width or thickness, is one approach to phase match pump, signal, and idler waves and target desired wavelengths. While such designs nominally support ultra-broadband wavelength conversion across spans that can exceed 100s of THz \cite{lu_visible_2020,domeneguetti_parametric_2021} about a fixed pump frequency, microresonator OPOs must also provide high-power and continuously single-mode operation with customizable outputs to be ideal converters.

To achieve high output power, one might first consider simply increasing the pump power to a microresonator OPO, anticipating a corresponding increase of signal and idler powers. In the high-pump-power regime, however, additional nonlinear interactions such as cluster comb formation and nondegenerate FWM activate and limit conversion to desired waves and degrade single mode operation \cite{sayson_2018_cluster,matsko_clustered_2016,stone2022conversion,stone2022efficient}. This behavior arises due to the mode structure of conventional microresonators that preserve free-spectral-range (FSR); resonances that neighbor the primary OPO signal and idler modes are themselves nearly phase-matched through nondegenerate FWM, thus supporting parametric gain at sufficiently high pump power. Analogous to lasers, where gratings control optical feedback to evolve multi-longitudinal-mode devices into high-power, single-mode sources, microresonator OPO phase matching can in principle be optimized for continuous operation versus pump power to realize practical sources.

\begin{figure*}[t] 
\centering \includegraphics[width=0.85\textwidth]{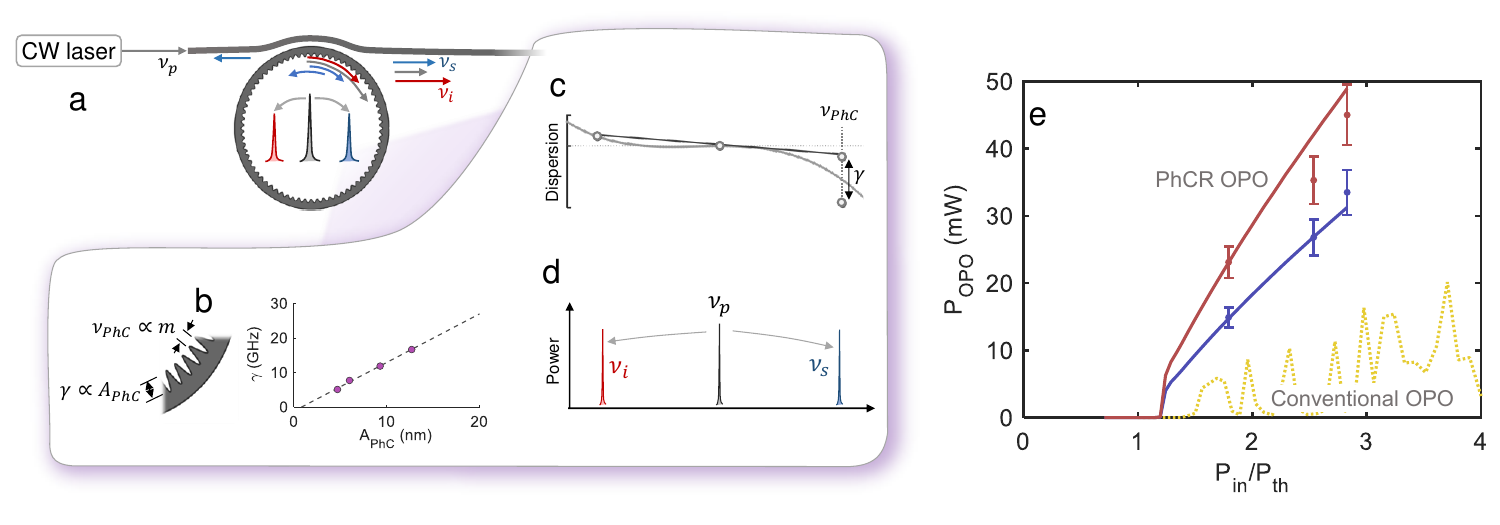}
\caption{(a) Schematic of a nanophotonic oscillator for high power wavelength conversion. (b) Parameters that select photonic crystal (PhC) frequency, $\nu_{PhC}$, via mode $m$ and PhC magnitude, $\gamma$, through amplitude $A_{PhC}$. Experimental (circles) and linear trend between $A_{PhC}$ and $\gamma$ in small bandgap regime. (c) Integrated dispersion with a line connecting resonant modes that satisfy frequency and phase matching. (d) Forward output spectrum showing pump laser with frequency $\nu_p$ converted to signal ($\nu_s$) and idler ($\nu_i$) waves. (e) On-chip power of converted waves. Modeled signal (blue line) and idler (red line) powers are in good agreement with experimental signal (blue dots) and idler (red dots). Modeled power conversion in a conventional microresonator OPO (yellow dashed).}

\label{fig1}
\end{figure*}

Recently, controlling such interactions has been explored in photonic-crystal resonators (PhCRs) \cite{lu2014selective,yu2021spontaneous,stone_wavelength-accurate_2023,lu_kerr_2022,black_optical-parametric_2022,moille_fourier_2023,liu2024threshold,brodnik_nanopatterned_2025} that lend programmable phase matching on a mode-by-mode basis. Through lithographically defined periodic sidewall modulations, such devices use coherent backscattering to couple clockwise (CW) and counterclockwise (CCW) propagating fields and lift the degeneracy of modes that satisfy the Bragg condition. The resulting non-degenerate, standing-wave modes are shifted to a higher (blue-shifted) and lower (red-shifted) frequency from the unperturbed resonance, proportional to the backscattering rate controlled by the photonic crystal (PhC) amplitude. This adjustable mode splitting in essence provides control of individual resonance frequencies, yielding a degree of freedom to impose phase matching independent of geometric dispersion \cite{yu2021spontaneous}. Devices with a single PhC splitting the pump mode address challenges associated with sensitive dispersion regimes \cite{brodnik_nanopatterned_2025} and enable operation in normal GVD, nominally unsupportive of direct FWM processes, to realize nearband OPO wavelength conversion \cite{black_optical-parametric_2022} and high efficiency soliton formation \cite{zang2024laserpower,10.1063/5.0191602}. Multi-PhC, bandgap-detuned devices, whereby bandgaps are designed on modes other than the pump mode, enable directional control and improved conversion efficiency of a broad suite of nonlinear processes \cite{jin_bandgap-detuned_2024}. Recently, a key advancement for wavelength-selective OPOs using PhCs on only signal or idler modes lends phase matching selectivity for wavelength accuracy and continuously tunable converted light \cite{stone_wavelength-accurate_2023}.

Here, we explore a paradigm for high-power wavelength conversion that couples microresonator dispersion engineering with PhC-activated phase matching. First, we shape dispersion using waveguide geometry to balance operation in the zero-GVD region, supportive of widespan (>10s of THz) OPO \cite{sayson_octave-spanning_2019}, with slight normal-GVD to broadly inhibit direct OPO phase matching. We then design a PhC targeting an OPO signal mode to bring the blue-shifted resonance back into phase matching while neighboring modes remain detuned, thus activating only the desired, single FWM conversion process. These nanophotonic oscillators unlock continuously single-mode operation at high power which we demonstrate by generating more than 40 mW in the idler wave and more than 30 mW in the combined forward and backward signal wave while maintaining >40 dB side mode suppression ratio (SMSR). We develop a modified Lugiato-Lefever-equation (LLE) \cite{chembo_spatiotemporal_2013} model that captures the PhC dynamics to directly compare nanophotonic oscillator performance to conventional microresonator OPOs. We demonstrate the broad spectral access of this approach in devices with nominally identical geometric GVD that use PhCs to activate four different signal modes for high-power conversion. Our experiments demonstrate nonlinear selectivity for continuously-single-mode operation of high-power wavelength converters for future coherent sources across the visible and near-IR.

\vspace{5mm}
\noindent\textbf{Phase matching and nonlinear selectivity}
\vspace{1mm}

We present nanophotonic oscillators with signal-mode PhC activation in Fig. 1 and further demonstrate high-power operation with input-output performance in agreement with modeling. We fabricate devices on the tantala nonlinear photonics platform \cite{black_group-velocity-dispersion_2021,jung_tantala_2021}, which offers low optical loss, high Kerr nonlinearity, low thermo-optic coefficient, and low temperature processing requirements. Figure 1a shows a nanophotonic oscillator pumped by a laser at frequency $\nu_p$, generating a forward-propagating idler wave ($\nu_i)$ and forward- and backward-propagating signal waves ($\nu_s$). In Fig. 1b, we illustrate the PhC design with amplitude parameter, $A_\text{PhC}$, and azimuthal mode selection, $m$, that control mode split magnitude, $\gamma$, and PhC frequency, $\nu_\text{PhC}$, respectively. In a representative device, we measure $\gamma$ for various $A_\text{PhC}$ values to show robust control over mode splitting for designable phase matching. Figure 1c shows a representative integrated dispersion ($D_{\text{int}}$) curve \cite{fujii2020dispersion}. Phase matching is displayed as a straight line connecting the participating signal and idler modes symmetric about the pump \cite{black2023nonlinear}. In this framework, phase matching is satisfied for only the PhC-activated, blue-shifted signal mode.

We design nanophotonic oscillators by first modeling geometric GVD with a finite element solver. For a given microresonator GVD, we calculate the FWM frequency error, $\delta_\nu=2\nu_p-\nu_s-\nu_i$ between resonant frequencies for pump, signal, and idler mode numbers $m_p$, $m_s$, and $m_i$ ($m_s$ and $m_i$ symmetric about $m_p$ \cite{kippenberg_kerr-nonlinearity_2004}) and frequencies $\nu_p$, $\nu_s$, and $\nu_i$. The frequency error gives the shift required to phase match a given FWM process; designing a PhC with $\gamma = 2\delta\nu$ on mode $m_s$ compensates the GVD-imposed detuning for only this selected mode, thereby rebalancing phase matching for the desired process. The characteristic forward-propagating output waves are shown in the optical spectrum in Fig. 1d, and we note that the signal wave power is distributed between forward and backward waves by the PhC. 

To explore complex nonlinear dynamics and aid in optimizing PhC designs for high output power operation, we develop a modified LLE model that expands other frequency-domain, split-step implementations by including coupling between clockwise (CW) and counter-clockwise (CCW) fields to account for PhC interactions \cite{liu2024threshold}. Our LLE model is given here, as

\begin{equation}\label{eq:LLE}
\begin{split}
    \frac{dE\fwd_\mu}{dt} = -(1 + i(\alpha + D_{\text{int}}))E\fwd_\mu \\ 
     + i(\sum_{\mu_1, \mu_2} E_{\mu_1}\fwd E\fwd_{\mu_2}E\fwd_{(\mu_1 + \mu_2 - \mu)} + 2E\fwd_\mu \sum_{\mu_3}|E\bck_{\mu_3}|^2) \\
     + F\delta_{\mu, 0} + i\frac{\gamma(\mu)}{2}E\bck_\mu,
\end{split}
\end{equation}

\noindent where $E\fwd_\mu$ ($E\bck_\mu$) is the  CW (CCW) field of the $\mu^{\text{th}}$ mode relative to the pump, $\alpha$ is the pump frequency detuning \cite{stone2022conversion}, $F$ is the driving field, and $\gamma(\mu)$ is the PhC-induced bandgap at relative mode $\mu$. $D_{\text{int}}$ is the integrated dispersion \cite{fujii2020dispersion}, defined as 

\begin{equation} \label{dint_polynomial}
\begin{split}
 \omega(\mu) & =  \omega_{0} + {D_1}\mu + \frac{D_2\mu^2}{2} + \frac{D_3\mu^3}{6} + ... \\
 & = {\omega_0} + {D_1}\mu + D_\textrm{int}.
\end{split}
\end{equation} 
 
\noindent Electric field quantities are normalized to the pump driving field at threshold, and frequency quantities are normalized to resonator halfwidths ($\Delta\nu/2 = \nu_0/(2Q_L)$, where $Q_L$ is the loaded quality factor). The model assumes uniform critical coupling of all resonator modes.

To consider phase matching due to detuning and power dependent Kerr self- and cross-phase modulation (SPM, XPM), we use the LLE to analyze and fine-tune $\gamma$ for high power. Based on these analyses, we show that geometries that realize weakly normal GVD sufficiently suppress nonlinear interactions while still affording widespan, PhC-activated phase matching with modest $A_{PhC}$.

We make the connection between design, LLE simulation, and experiment by operating a nanophotonic oscillator and show high-power operation in agreement to modeling in Fig. 1e. We plot our highest experimentally measured output ($\nu_i \approx 1003$ nm, $\nu_s \approx 928$ nm) power as a function of pump ($\nu_p \approx 964$ nm) power, measuring 41 mW of forward-propagating idler and 30 mW of total forward- and backward-propagating signal.  We present simulated output power using Eq.~\ref{eq:LLE} that agrees well with the measured powers. These simulation results include mode-dependent coupling parameters for the characterized device, discussed in later sections.  On the same plot, we also model the performance of a conventional zero-GVD microresonator OPO targeting the same output waves using geometric GVD alone to realize phase matching. From this model, we see that broadband parametric gain supported in conventional OPO designs results in limited conversion efficiency due to nonlinear competition and mode-hopping as power is increased. In contrast, modeled and experimental nanophotonic oscillators demonstrate continuously single-mode operation and monotonically increasing output power well into the high-pump-power regime. 

\vspace{5mm}
\noindent\textbf{Power dependent nonlinear dynamics}
\vspace{1mm}

\begin{figure*}[ht] \centering 
\includegraphics[width=0.85\textwidth]{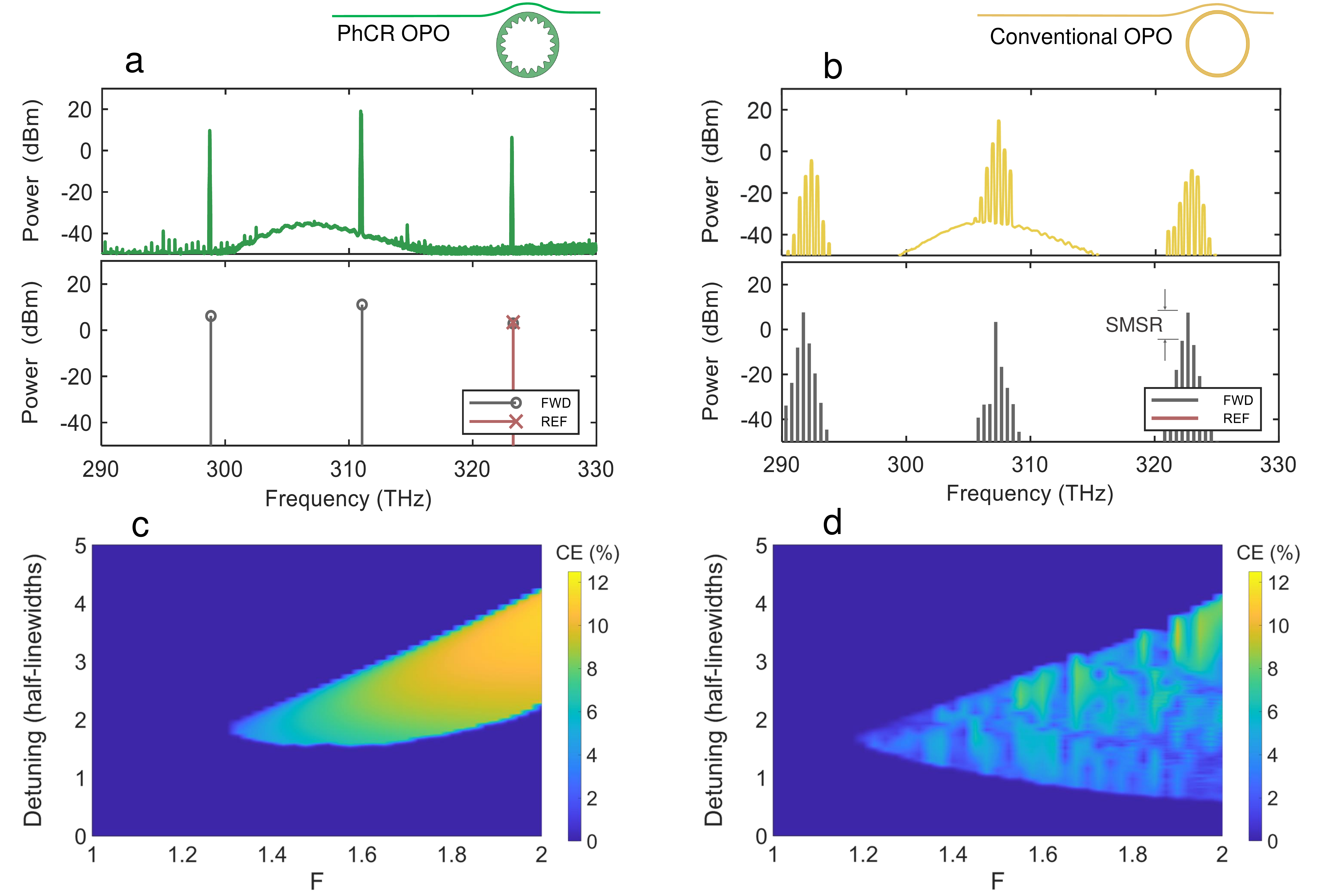}
\caption{(a,b) Experimental (top) and simulated (bottom) OPO spectra of a nanophotonic oscillator and conventional microresonator OPO, respectively. (c,d) Conversion efficiency as a function of pump-detuning and normalized driving field for a nanophotonic oscillator and conventional microresonator OPO, respectively.
}
\label{fig2}
\end{figure*}

\noindent With an experimental demonstration of nanophotonic oscillators, according to our design approach and models, we next delve into the dynamics that govern performance. At threshold, OPO processes are seeded by vacuum fluctuations. Above threshold, these processes evolve according to the resonator dispersion, pump power, and pump frequency detuning. For a given device to be practical as a wavelength converter, its evolution through changes in power and detuning should be well-behaved. We employ the LLE model to observe this evolution over the phase space of pump power and detuning for both nanophotonic oscillators and conventional resonators.

Figure 2 compares the nonlinear dynamics of nanophotonic oscillators to those of conventional microresonator OPOs in the high power regime. In particular, we show nanophotonic oscillators to have well-behaved input/output power relations and continuously single-mode operation in contrast to conventional OPOs that seed nonlinear competition and exhibit mode hopping. In Fig. 2a, we show experimental and LLE modeled optical spectra of a representative nanophotonic oscillator operating at high pump power. The nanophotonic oscillator maintains SMSR exceeding 40 dB and approaches coupling-limited conversion efficiency \cite{sayson_octave-spanning_2019} by maintaining phase matching between only the desired signal mode to the pump and idler by the PhC design. Figure 2b compares experimental and modeled optical spectra for the waves in a conventional microresonator OPO. Conventional OPO conversion efficiency saturates and SMSR remains low due to nondegenerate FWM between modes other than the primary phase matched signal and idler waves.

Given the confirmation between experimentally measured spectra and LLE modeling, we further use the LLE to examine conversion efficiency across the pump power and detuning phase space for nanophotonic oscillators and conventional microresonator OPOs in Fig. 2 (c) and (d), respectively. By sweeping over the pump detuning space, we capture the effects of SPM and XPM and account for the resulting changes in phase matching and intracavity power. The region of the power phase space spans powers below, at, and several times above threshold. For devices with uniform critical coupling, we calculate the total conversion efficiency in terms of total photon flux \cite{stone2022conversion} in the signal wave using the fields modeled in the LLE. For nanophotonic oscillators, this output wave is the PhC-activated signal mode across the full parameter sweep, and we sum the power in the forwards- and backwards- propagating fields. For conventional microresonator OPOs that, in general, have a changing maximum-power output wave due to mode hopping, we consider the conversion efficiency for the highest power signal wave at any given set of power and detuning. Our modeling shows a monotonic increase in converted power for nanophotonic oscillators. In contrast, conventional designs exhibit chaotic dynamics due to nonlinear competition and the activation of additional FWM processes supported by broad parametric gain. Further, nonlinear states in conventional microresonators show a high sensitivity to detuning.

\vspace{5mm}
\noindent\textbf{Wavelength versatility and power}
\vspace{1mm}

\begin{figure*}[ht!!] \centering \includegraphics[width=0.85\textwidth]{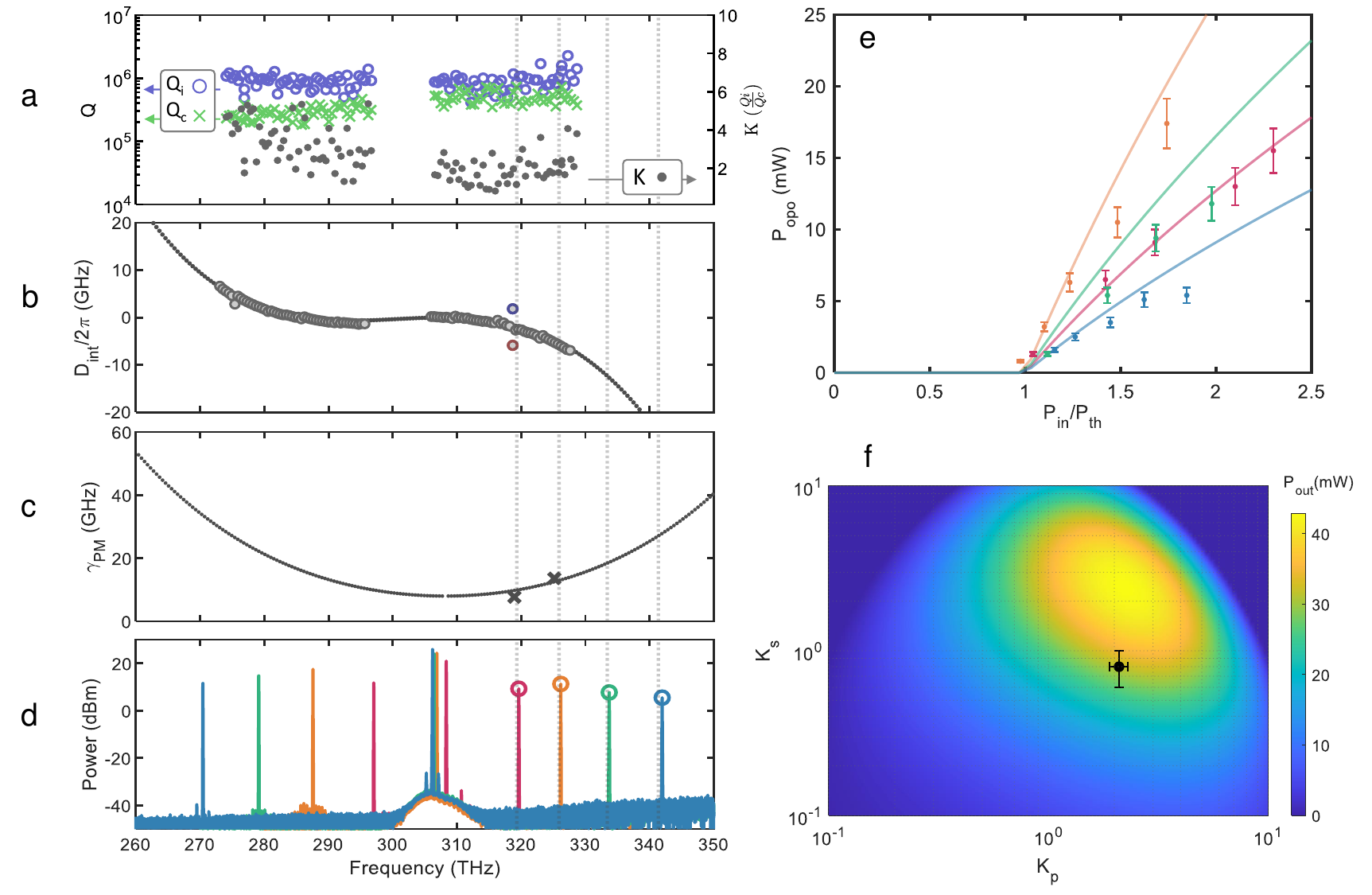}
\caption{(a) Measured $Q_i$ (blue circles) and $Q_c$ (green crosses), and associated coupling constant K. (b) Integrated dispersion of a representative fabricated PhCR designed for wavelength conversion to 940nm.  Gray circles are measured values and black dots are values obtained from finite-element simulations.  The PhC-induced mode-split is shown with blue and red circles for the higher and lower frequency standing-wave modes, respectively. (c) The bandgap needed to satisfy the phase matching condition, $\gamma_{\text{PM}}$, for various target signal frequencies at high power. The dotted line shows the predicted necessary bandgap while the x's show the measured bandgap on operational devices. (d) Experimental output spectra for PhCRs with targets of 940, 920, 900, and 880 nm. (e) Total output power as a function of input power (normalized to threshold), colored corresponding to the devices shown in panel (d).  (f) Theoretical output power as a function of coupling for the signal ($K_s$) and pump ($K_p$) modes with a fixed pump power, chosen here as 220 mW. The experimental data point corresponds to the device shown in Fig. 1e.}
\label{fig3}
\end{figure*}

\noindent
Since a key use of nanophotonic oscillators is to provide high-power light sources customized for user wavelengths, we next demonstrate wavelength versatility. Using the same pump laser, we design PhCs to activate four different output waves across four different devices. These devices share the same nominal waveguide width, $w$, and thickness, $t$, and are designed following our paradigm in the weakly normal GVD that balances widespan phase matching and nonlinear inhibition. Each devive uses a different PhC parameter pair, $m_{PhC}$ and $A_{PhC}$, to set the OPO signal mode and bandgap magnitude, respectively, and control which single set of FWM modes are phase-matched for pump conversion to targeted output waves. We characterize this design space and other important considerations in Fig. 3, measuring GVD, quality factor (Q), coupling, optical spectra, and output power for representative devices.

We use two widely tunable lasers in the 980 nm and 1064 nm bands to characterize this set of nanophotonic oscillators with $w = 930~\text{nm}$ and $t = 570~\text{nm}$. For a single device from this set, Fig. 3a shows representative intrinsic ($Q_i$) and coupling ($Q_c$) quality factors and the associated coupling constant, $K=Q_i/Q_c$, which mediates nonlinear threshold and slope efficiency. On the same device, we measure GVD in Fig. 3b and label the PhC bandgap that we designed on the signal mode. For these devices, phase matching is enforced between the blue-shifted (blue circle) signal mode and corresponding pump and idler modes. In Fig. 3c, we use the cold-cavity frequency error, $\delta_\nu$, with an assumed effective pump resonance shift of 2 GHz from SPM and XPM to show the $\gamma$ needed to activate phase matching for different output waves. This, and LLE modeling of hot-cavity phase matching along with measurements of $\gamma$ vs $A_{PhC}$ define the center value of parameter sweeps of $A_{PhC}$ in fabrication. We pump the four devices in this design group and record the observed output spectra in Fig. 3d.  

We next complement the analysis of high-power operation given by our modified LLE with an analytic approximation assuming fixed, optimum phase matching, of conversion efficiency and power that includes mode-dependent coupling \cite{perez_high_performance_2023}. The theoretical maximum conversion efficiency to an output signal wave in a conventional OPO depends on the coupling constant of the pump ($K_p$) and signal mode ($K_s$) as
    
\begin{equation}\label{maxCE}
\text{CE}_\text{max} = \frac{\nu_s}{2\nu_p}\frac{K_s K_p}{(K_s + 1)(K_p + 1)},
\end{equation}

\noindent where $\nu_s$ and $\nu_p$ are the signal and pump wave frequencies, respectively, to represent $\text{CE}_\text{max}$ in terms of optical power rather than photon flux.
This result suggests operating in the overcoupled regime ($K > 1$) to maximize conversion efficiency. While true, overcoupling increases OPO threshold in a uniformly coupled device \cite{lu_visible_2020,briles_generating_2020} according to

\begin{equation}\label{threshold}
P_\text{th} \propto \frac{(K+1)^3}{K}.
\end{equation}

\noindent We modify Eq. 4 to account for mode-dependent coupling below, as,

\begin{equation}\label{threshold}
P_\text{th} \propto \frac{(K_p+1)^2\sqrt{(K_s+1)(K_i+1)}}{K_p}.
\end{equation}

\noindent We make Eqs. 3 and 5 applicable to our nanophotonic oscillators, where the signal mode is split by a PhC interaction, by making the substitution $(K_s+1) \rightarrow 2~(K_s+1)$. This factor of 2 arises due to FWM occurring only in the forward (CW) direction, co-propogating the pump and idler waves, while the signal field has PhC-coupled forward and backward components.

Furthermore, conversion efficiency also depends on the factor above OPO threshold at which the pump laser is operated, saturating at $\text{CE} = \text{CE}_\text{max}$ when $P_{\text{in}} \geq 4P_\text{th}$, according to
    
\begin{equation}\label{observedCE}
\text{CE} = \text{CE}_\text{max} \frac{4}{X} (\sqrt{X} - 1),
\end{equation}

\noindent where $X = P_{\text{in}}/P_\text{th}$ and $1 \leq X \leq 4$ \cite{sayson_octave-spanning_2019}. From Eqs.~\ref{maxCE}-\ref{observedCE}, we see that to maximize CE, a nanophotonic oscillator should be overcoupled to increase the outcoupling rate and maximum CE, but not so much as to significantly increase threshold.

In Fig. 3e, we plot the total measured signal power (summed forward and backward waves) of these four devices as a function of pump power in colors corresponding to the output spectra of Fig. 3d. We use the measured coupling and threshold of the tested devices in Eqs.~\ref{maxCE}-\ref{observedCE} to produce the analytical curves shown in the same figure, which represents the coupling-dependent maximum output power with the assumption of fixed, optimal phase matching.

By combining the variable-coupling analytical approximations with our critical-coupling LLE model, we can explore operation of practical devices considering a fixed pump power. The interplay of conversion efficiency and threshold power under such conditions is shown in Fig.~\ref{fig3}f for a fixed $K_i$ and variable $K_p$ and $K_s$. This relation exhibits a maximum output power at specific coupling values for an ideal device. The experimental data point in Fig.~\ref{fig3}f corresponds to the coupling of the device whose output power is shown in Fig.~\ref{fig1}e. In experiment, we estimate  the maximum on-chip pump power after chip facet coupling losses to be 220 mW. The tested device has estimated couplings of $K_p = 2.1$, $K_s = 0.8$, and $K_i = 2.3$ with a threshold of $P_{\text{th}} = 80 \pm 20$ mW. Inserting these values into our mode-dependent optimum phase matching expression suggests a possible improvement in output signal power had the device been designed with larger $K_p$, $K_i$, and, particularly, $K_s$.

\vspace{5mm}
\noindent\textbf{Conclusion} 
\vspace{1mm} \newline
\noindent
We have explored a high-power wavelength conversion paradigm that couples GVD engineering with nanophotonic phase matching to realize high-power, spectrally pure wavelength conversion. By using a photonic crystal to enforce phase matching between only a single set of chosen FWM modes, other nonlinear interactions that limit conventional microresonator OPOs operating at high-power are inhibited. We show this control experimentally and with LLE models for several nanophotonic oscillators designed for a range of wavelength conversion targets. Measured devices demonstrate well-behaved input-output dynamics and consistently high SMSR, contrasting conventional microresonator OPOs with performance typically limited by nonlinear competition. These results show that nanophotonic oscillators are an evolution of integrated wavelength converters in terms of power, robustness in operation, and scalability to support chip-scale visible and near-IR optical technologies.

\vspace{5mm}
\noindent\textbf{Acknowledgments.} We thank Nitesh Chauhan and Yan Jin for careful review of the manuscript.

\vspace{2mm}

\noindent\textbf{Funding.} This research has been funded by the DARPA LUMOS program HR0011-20-2-0046, AFOSR FA9550-20-1-0004 Project Number 19RT1019, NSF Quantum Leap Challenge Institute Award OMA - 2016244, and NIST.

\vspace{2mm}

\noindent\textbf{Disclosures.} The authors declare no competing interests. This work is a contribution of the US Government and is not subject to US copyright. Mention of specific companies or trade names is for scientific communication only and does not constitute an endorsement by NIST. 

\vspace{2mm}

\noindent\textbf{Data Availability Statement.}
The data that support the plots within this paper and other findings of this study are available from the corresponding author upon reasonable request.

\clearpage

\bibliography{BIBLIOGRAPHY_HP_nanophotonic_oscillators} 

\begin{thebibliography}{32}%
\makeatletter
\providecommand \@ifxundefined [1]{%
 \@ifx{#1\undefined}
}%
\providecommand \@ifnum [1]{%
 \ifnum #1\expandafter \@firstoftwo
 \else \expandafter \@secondoftwo
 \fi
}%
\providecommand \@ifx [1]{%
 \ifx #1\expandafter \@firstoftwo
 \else \expandafter \@secondoftwo
 \fi
}%
\providecommand \natexlab [1]{#1}%
\providecommand \enquote  [1]{``#1''}%
\providecommand \bibnamefont  [1]{#1}%
\providecommand \bibfnamefont [1]{#1}%
\providecommand \citenamefont [1]{#1}%
\providecommand \href@noop [0]{\@secondoftwo}%
\providecommand \href [0]{\begingroup \@sanitize@url \@href}%
\providecommand \@href[1]{\@@startlink{#1}\@@href}%
\providecommand \@@href[1]{\endgroup#1\@@endlink}%
\providecommand \@sanitize@url [0]{\catcode `\\12\catcode `\$12\catcode `\&12\catcode `\#12\catcode `\^12\catcode `\_12\catcode `\%12\relax}%
\providecommand \@@startlink[1]{}%
\providecommand \@@endlink[0]{}%
\providecommand \url  [0]{\begingroup\@sanitize@url \@url }%
\providecommand \@url [1]{\endgroup\@href {#1}{\urlprefix }}%
\providecommand \urlprefix  [0]{URL }%
\providecommand \Eprint [0]{\href }%
\providecommand \doibase [0]{https://doi.org/}%
\providecommand \selectlanguage [0]{\@gobble}%
\providecommand \bibinfo  [0]{\@secondoftwo}%
\providecommand \bibfield  [0]{\@secondoftwo}%
\providecommand \translation [1]{[#1]}%
\providecommand \BibitemOpen [0]{}%
\providecommand \bibitemStop [0]{}%
\providecommand \bibitemNoStop [0]{.\EOS\space}%
\providecommand \EOS [0]{\spacefactor3000\relax}%
\providecommand \BibitemShut  [1]{\csname bibitem#1\endcsname}%
\let\auto@bib@innerbib\@empty
\bibitem [{\citenamefont {Lu}\ \emph {et~al.}(2019)\citenamefont {Lu}, \citenamefont {Li}, \citenamefont {Westly}, \citenamefont {Moille}, \citenamefont {Singh}, \citenamefont {Anant},\ and\ \citenamefont {Srinivasan}}]{lu_chip-integrated_2019}%
  \BibitemOpen
  \bibfield  {author} {\bibinfo {author} {\bibfnamefont {X.}~\bibnamefont {Lu}}, \bibinfo {author} {\bibfnamefont {Q.}~\bibnamefont {Li}}, \bibinfo {author} {\bibfnamefont {D.~A.}\ \bibnamefont {Westly}}, \bibinfo {author} {\bibfnamefont {G.}~\bibnamefont {Moille}}, \bibinfo {author} {\bibfnamefont {A.}~\bibnamefont {Singh}}, \bibinfo {author} {\bibfnamefont {V.}~\bibnamefont {Anant}},\ and\ \bibinfo {author} {\bibfnamefont {K.}~\bibnamefont {Srinivasan}},\ }\bibfield  {title} {\bibinfo {title} {Chip-integrated visible–telecom entangled photon pair source for quantum communication},\ }\href {https://doi.org/10.1038/s41567-018-0394-3} {\bibfield  {journal} {\bibinfo  {journal} {Nature Physics}\ }\textbf {\bibinfo {volume} {15}},\ \bibinfo {pages} {373} (\bibinfo {year} {2019})}\BibitemShut {NoStop}%
\bibitem [{\citenamefont {Yang}\ \emph {et~al.}(2021)\citenamefont {Yang}, \citenamefont {Jahanbozorgi}, \citenamefont {Jeong}, \citenamefont {Sun}, \citenamefont {Pfister}, \citenamefont {Lee},\ and\ \citenamefont {Yi}}]{Yang_2021}%
  \BibitemOpen
  \bibfield  {author} {\bibinfo {author} {\bibfnamefont {Z.}~\bibnamefont {Yang}}, \bibinfo {author} {\bibfnamefont {M.}~\bibnamefont {Jahanbozorgi}}, \bibinfo {author} {\bibfnamefont {D.}~\bibnamefont {Jeong}}, \bibinfo {author} {\bibfnamefont {S.}~\bibnamefont {Sun}}, \bibinfo {author} {\bibfnamefont {O.}~\bibnamefont {Pfister}}, \bibinfo {author} {\bibfnamefont {H.}~\bibnamefont {Lee}},\ and\ \bibinfo {author} {\bibfnamefont {X.}~\bibnamefont {Yi}},\ }\bibfield  {title} {\bibinfo {title} {A squeezed quantum microcomb on a chip},\ }\bibfield  {journal} {\bibinfo  {journal} {Nature Communications}\ }\textbf {\bibinfo {volume} {12}},\ \href {https://doi.org/10.1038/s41467-021-25054-z} {10.1038/s41467-021-25054-z} (\bibinfo {year} {2021})\BibitemShut {NoStop}%
\bibitem [{\citenamefont {Del’Haye}\ \emph {et~al.}(2007)\citenamefont {Del’Haye}, \citenamefont {Schliesser}, \citenamefont {Arcizet}, \citenamefont {Wilken}, \citenamefont {Holzwarth},\ and\ \citenamefont {Kippenberg}}]{delhaye_optical_2007}%
  \BibitemOpen
  \bibfield  {author} {\bibinfo {author} {\bibfnamefont {P.}~\bibnamefont {Del’Haye}}, \bibinfo {author} {\bibfnamefont {A.}~\bibnamefont {Schliesser}}, \bibinfo {author} {\bibfnamefont {O.}~\bibnamefont {Arcizet}}, \bibinfo {author} {\bibfnamefont {T.}~\bibnamefont {Wilken}}, \bibinfo {author} {\bibfnamefont {R.}~\bibnamefont {Holzwarth}},\ and\ \bibinfo {author} {\bibfnamefont {T.~J.}\ \bibnamefont {Kippenberg}},\ }\bibfield  {title} {\bibinfo {title} {Optical frequency comb generation from a monolithic microresonator},\ }\href {https://doi.org/10.1038/nature06401} {\bibfield  {journal} {\bibinfo  {journal} {Nature}\ }\textbf {\bibinfo {volume} {450}},\ \bibinfo {pages} {1214} (\bibinfo {year} {2007})}\BibitemShut {NoStop}%
\bibitem [{\citenamefont {Pasquazi}\ \emph {et~al.}(2018)\citenamefont {Pasquazi}, \citenamefont {Peccianti}, \citenamefont {Razzari}, \citenamefont {Moss}, \citenamefont {Coen}, \citenamefont {Erkintalo}, \citenamefont {Chembo}, \citenamefont {Hansson}, \citenamefont {Wabnitz}, \citenamefont {Del’Haye}, \citenamefont {Xue}, \citenamefont {Weiner},\ and\ \citenamefont {Morandotti}}]{pasquazi_micro-combs_2018}%
  \BibitemOpen
  \bibfield  {author} {\bibinfo {author} {\bibfnamefont {A.}~\bibnamefont {Pasquazi}}, \bibinfo {author} {\bibfnamefont {M.}~\bibnamefont {Peccianti}}, \bibinfo {author} {\bibfnamefont {L.}~\bibnamefont {Razzari}}, \bibinfo {author} {\bibfnamefont {D.~J.}\ \bibnamefont {Moss}}, \bibinfo {author} {\bibfnamefont {S.}~\bibnamefont {Coen}}, \bibinfo {author} {\bibfnamefont {M.}~\bibnamefont {Erkintalo}}, \bibinfo {author} {\bibfnamefont {Y.~K.}\ \bibnamefont {Chembo}}, \bibinfo {author} {\bibfnamefont {T.}~\bibnamefont {Hansson}}, \bibinfo {author} {\bibfnamefont {S.}~\bibnamefont {Wabnitz}}, \bibinfo {author} {\bibfnamefont {P.}~\bibnamefont {Del’Haye}}, \bibinfo {author} {\bibfnamefont {X.}~\bibnamefont {Xue}}, \bibinfo {author} {\bibfnamefont {A.~M.}\ \bibnamefont {Weiner}},\ and\ \bibinfo {author} {\bibfnamefont {R.}~\bibnamefont {Morandotti}},\ }\bibfield  {title} {\bibinfo {title} {Micro-combs: {A} novel generation of optical sources},\ }\href {https://doi.org/10.1016/j.physrep.2017.08.004} {\bibfield
  {journal} {\bibinfo  {journal} {Physics Reports}\ }\bibinfo {series} {Micro-combs: {A} novel generation of optical sources},\ \textbf {\bibinfo {volume} {729}},\ \bibinfo {pages} {1} (\bibinfo {year} {2018})}\BibitemShut {NoStop}%
\bibitem [{\citenamefont {Herr}\ \emph {et~al.}(2014)\citenamefont {Herr}, \citenamefont {Brasch}, \citenamefont {Jost}, \citenamefont {Wang}, \citenamefont {Kondratiev}, \citenamefont {Gorodetsky},\ and\ \citenamefont {Kippenberg}}]{herr_temporal_2014}%
  \BibitemOpen
  \bibfield  {author} {\bibinfo {author} {\bibfnamefont {T.}~\bibnamefont {Herr}}, \bibinfo {author} {\bibfnamefont {V.}~\bibnamefont {Brasch}}, \bibinfo {author} {\bibfnamefont {J.~D.}\ \bibnamefont {Jost}}, \bibinfo {author} {\bibfnamefont {C.~Y.}\ \bibnamefont {Wang}}, \bibinfo {author} {\bibfnamefont {N.~M.}\ \bibnamefont {Kondratiev}}, \bibinfo {author} {\bibfnamefont {M.~L.}\ \bibnamefont {Gorodetsky}},\ and\ \bibinfo {author} {\bibfnamefont {T.~J.}\ \bibnamefont {Kippenberg}},\ }\bibfield  {title} {\bibinfo {title} {Temporal solitons in optical microresonators},\ }\href {https://doi.org/10.1038/nphoton.2013.343} {\bibfield  {journal} {\bibinfo  {journal} {Nature Photonics}\ }\textbf {\bibinfo {volume} {8}},\ \bibinfo {pages} {145} (\bibinfo {year} {2014})},\ \bibinfo {note} {publisher: Nature Publishing Group}\BibitemShut {NoStop}%
\bibitem [{\citenamefont {Lu}\ \emph {et~al.}(2020)\citenamefont {Lu}, \citenamefont {Moille}, \citenamefont {Rao}, \citenamefont {Westly},\ and\ \citenamefont {Srinivasan}}]{lu_visible_2020}%
  \BibitemOpen
  \bibfield  {author} {\bibinfo {author} {\bibfnamefont {X.}~\bibnamefont {Lu}}, \bibinfo {author} {\bibfnamefont {G.}~\bibnamefont {Moille}}, \bibinfo {author} {\bibfnamefont {A.}~\bibnamefont {Rao}}, \bibinfo {author} {\bibfnamefont {D.~A.}\ \bibnamefont {Westly}},\ and\ \bibinfo {author} {\bibfnamefont {K.}~\bibnamefont {Srinivasan}},\ }\bibfield  {title} {\bibinfo {title} {On-chip optical parametric oscillation into the visible: generating red, orange, yellow, and green from a near-infrared pump},\ }\href {https://doi.org/10.1364/OPTICA.393810} {\bibfield  {journal} {\bibinfo  {journal} {Optica}\ }\textbf {\bibinfo {volume} {7}},\ \bibinfo {pages} {1417} (\bibinfo {year} {2020})}\BibitemShut {NoStop}%
\bibitem [{\citenamefont {Domeneguetti}\ \emph {et~al.}(2021)\citenamefont {Domeneguetti}, \citenamefont {Zhao}, \citenamefont {Ji}, \citenamefont {Martinelli}, \citenamefont {Lipson}, \citenamefont {Gaeta},\ and\ \citenamefont {Nussenzveig}}]{domeneguetti_parametric_2021}%
  \BibitemOpen
  \bibfield  {author} {\bibinfo {author} {\bibfnamefont {R.~R.}\ \bibnamefont {Domeneguetti}}, \bibinfo {author} {\bibfnamefont {Y.}~\bibnamefont {Zhao}}, \bibinfo {author} {\bibfnamefont {X.}~\bibnamefont {Ji}}, \bibinfo {author} {\bibfnamefont {M.}~\bibnamefont {Martinelli}}, \bibinfo {author} {\bibfnamefont {M.}~\bibnamefont {Lipson}}, \bibinfo {author} {\bibfnamefont {A.~L.}\ \bibnamefont {Gaeta}},\ and\ \bibinfo {author} {\bibfnamefont {P.}~\bibnamefont {Nussenzveig}},\ }\bibfield  {title} {\bibinfo {title} {Parametric sideband generation in {CMOS}-compatible oscillators from visible to telecom wavelengths},\ }\href {https://doi.org/10.1364/OPTICA.404755} {\bibfield  {journal} {\bibinfo  {journal} {Optica}\ }\textbf {\bibinfo {volume} {8}},\ \bibinfo {pages} {316} (\bibinfo {year} {2021})}\BibitemShut {NoStop}%
\bibitem [{\citenamefont {Sun}\ \emph {et~al.}(2024)\citenamefont {Sun}, \citenamefont {Stone}, \citenamefont {Lu}, \citenamefont {Zhou}, \citenamefont {Song}, \citenamefont {Shi},\ and\ \citenamefont {Srinivasan}}]{sun_advancing_2024}%
  \BibitemOpen
  \bibfield  {author} {\bibinfo {author} {\bibfnamefont {Y.}~\bibnamefont {Sun}}, \bibinfo {author} {\bibfnamefont {J.}~\bibnamefont {Stone}}, \bibinfo {author} {\bibfnamefont {X.}~\bibnamefont {Lu}}, \bibinfo {author} {\bibfnamefont {F.}~\bibnamefont {Zhou}}, \bibinfo {author} {\bibfnamefont {J.}~\bibnamefont {Song}}, \bibinfo {author} {\bibfnamefont {Z.}~\bibnamefont {Shi}},\ and\ \bibinfo {author} {\bibfnamefont {K.}~\bibnamefont {Srinivasan}},\ }\bibfield  {title} {\bibinfo {title} {Advancing on-chip {Kerr} optical parametric oscillation towards coherent applications covering the green gap},\ }\href {https://doi.org/10.1038/s41377-024-01534-x} {\bibfield  {journal} {\bibinfo  {journal} {Light: Science \& Applications}\ }\textbf {\bibinfo {volume} {13}},\ \bibinfo {pages} {201} (\bibinfo {year} {2024})},\ \bibinfo {note} {publisher: Nature Publishing Group}\BibitemShut {NoStop}%
\bibitem [{\citenamefont {Sayson}\ \emph {et~al.}(2018)\citenamefont {Sayson}, \citenamefont {Pham}, \citenamefont {Webb}, \citenamefont {Ng}, \citenamefont {Trainor}, \citenamefont {Schwefel}, \citenamefont {Coen}, \citenamefont {Erkintalo},\ and\ \citenamefont {Murdoch}}]{sayson_2018_cluster}%
  \BibitemOpen
  \bibfield  {author} {\bibinfo {author} {\bibfnamefont {N.~L.~B.}\ \bibnamefont {Sayson}}, \bibinfo {author} {\bibfnamefont {H.}~\bibnamefont {Pham}}, \bibinfo {author} {\bibfnamefont {K.~E.}\ \bibnamefont {Webb}}, \bibinfo {author} {\bibfnamefont {V.}~\bibnamefont {Ng}}, \bibinfo {author} {\bibfnamefont {L.~S.}\ \bibnamefont {Trainor}}, \bibinfo {author} {\bibfnamefont {H.~G.~L.}\ \bibnamefont {Schwefel}}, \bibinfo {author} {\bibfnamefont {S.}~\bibnamefont {Coen}}, \bibinfo {author} {\bibfnamefont {M.}~\bibnamefont {Erkintalo}},\ and\ \bibinfo {author} {\bibfnamefont {S.~G.}\ \bibnamefont {Murdoch}},\ }\bibfield  {title} {\bibinfo {title} {Origins of clustered frequency combs in kerr microresonators},\ }\href {https://doi.org/10.1364/OL.43.004180} {\bibfield  {journal} {\bibinfo  {journal} {Opt. Lett.}\ }\textbf {\bibinfo {volume} {43}},\ \bibinfo {pages} {4180} (\bibinfo {year} {2018})}\BibitemShut {NoStop}%
\bibitem [{\citenamefont {Matsko}\ \emph {et~al.}(2016)\citenamefont {Matsko}, \citenamefont {Savchenkov}, \citenamefont {Huang},\ and\ \citenamefont {Maleki}}]{matsko_clustered_2016}%
  \BibitemOpen
  \bibfield  {author} {\bibinfo {author} {\bibfnamefont {A.~B.}\ \bibnamefont {Matsko}}, \bibinfo {author} {\bibfnamefont {A.~A.}\ \bibnamefont {Savchenkov}}, \bibinfo {author} {\bibfnamefont {S.-W.}\ \bibnamefont {Huang}},\ and\ \bibinfo {author} {\bibfnamefont {L.}~\bibnamefont {Maleki}},\ }\bibfield  {title} {\bibinfo {title} {Clustered frequency comb},\ }\href {https://doi.org/10.1364/OL.41.005102} {\bibfield  {journal} {\bibinfo  {journal} {Optics Letters}\ }\textbf {\bibinfo {volume} {41}},\ \bibinfo {pages} {5102} (\bibinfo {year} {2016})},\ \bibinfo {note} {publisher: Optica Publishing Group}\BibitemShut {NoStop}%
\bibitem [{\citenamefont {Stone}\ \emph {et~al.}(2022{\natexlab{a}})\citenamefont {Stone}, \citenamefont {Moille}, \citenamefont {Lu},\ and\ \citenamefont {Srinivasan}}]{stone2022conversion}%
  \BibitemOpen
  \bibfield  {author} {\bibinfo {author} {\bibfnamefont {J.~R.}\ \bibnamefont {Stone}}, \bibinfo {author} {\bibfnamefont {G.}~\bibnamefont {Moille}}, \bibinfo {author} {\bibfnamefont {X.}~\bibnamefont {Lu}},\ and\ \bibinfo {author} {\bibfnamefont {K.}~\bibnamefont {Srinivasan}},\ }\bibfield  {title} {\bibinfo {title} {Conversion efficiency in kerr-microresonator optical parametric oscillators: From three modes to many modes},\ }\href@noop {} {\bibfield  {journal} {\bibinfo  {journal} {Physical Review Applied}\ }\textbf {\bibinfo {volume} {17}},\ \bibinfo {pages} {024038} (\bibinfo {year} {2022}{\natexlab{a}})}\BibitemShut {NoStop}%
\bibitem [{\citenamefont {Stone}\ \emph {et~al.}(2022{\natexlab{b}})\citenamefont {Stone}, \citenamefont {Lu}, \citenamefont {Moille},\ and\ \citenamefont {Srinivasan}}]{stone2022efficient}%
  \BibitemOpen
  \bibfield  {author} {\bibinfo {author} {\bibfnamefont {J.~R.}\ \bibnamefont {Stone}}, \bibinfo {author} {\bibfnamefont {X.}~\bibnamefont {Lu}}, \bibinfo {author} {\bibfnamefont {G.}~\bibnamefont {Moille}},\ and\ \bibinfo {author} {\bibfnamefont {K.}~\bibnamefont {Srinivasan}},\ }\bibfield  {title} {\bibinfo {title} {Efficient chip-based optical parametric oscillators from 590 to 1150 nm},\ }\href@noop {} {\bibfield  {journal} {\bibinfo  {journal} {APL Photonics}\ }\textbf {\bibinfo {volume} {7}} (\bibinfo {year} {2022}{\natexlab{b}})}\BibitemShut {NoStop}%
\bibitem [{\citenamefont {Lu}\ \emph {et~al.}(2014)\citenamefont {Lu}, \citenamefont {Rogers}, \citenamefont {Jiang},\ and\ \citenamefont {Lin}}]{lu2014selective}%
  \BibitemOpen
  \bibfield  {author} {\bibinfo {author} {\bibfnamefont {X.}~\bibnamefont {Lu}}, \bibinfo {author} {\bibfnamefont {S.}~\bibnamefont {Rogers}}, \bibinfo {author} {\bibfnamefont {W.~C.}\ \bibnamefont {Jiang}},\ and\ \bibinfo {author} {\bibfnamefont {Q.}~\bibnamefont {Lin}},\ }\bibfield  {title} {\bibinfo {title} {Selective engineering of cavity resonance for frequency matching in optical parametric processes},\ }\href@noop {} {\bibfield  {journal} {\bibinfo  {journal} {Applied Physics Letters}\ }\textbf {\bibinfo {volume} {105}},\ \bibinfo {pages} {151104} (\bibinfo {year} {2014})}\BibitemShut {NoStop}%
\bibitem [{\citenamefont {Yu}\ \emph {et~al.}(2021)\citenamefont {Yu}, \citenamefont {Cole}, \citenamefont {Jung}, \citenamefont {Moille}, \citenamefont {Srinivasan},\ and\ \citenamefont {Papp}}]{yu2021spontaneous}%
  \BibitemOpen
  \bibfield  {author} {\bibinfo {author} {\bibfnamefont {S.-P.}\ \bibnamefont {Yu}}, \bibinfo {author} {\bibfnamefont {D.~C.}\ \bibnamefont {Cole}}, \bibinfo {author} {\bibfnamefont {H.}~\bibnamefont {Jung}}, \bibinfo {author} {\bibfnamefont {G.~T.}\ \bibnamefont {Moille}}, \bibinfo {author} {\bibfnamefont {K.}~\bibnamefont {Srinivasan}},\ and\ \bibinfo {author} {\bibfnamefont {S.~B.}\ \bibnamefont {Papp}},\ }\bibfield  {title} {\bibinfo {title} {Spontaneous pulse formation in edgeless photonic crystal resonators},\ }\href@noop {} {\bibfield  {journal} {\bibinfo  {journal} {Nature Photonics}\ }\textbf {\bibinfo {volume} {15}},\ \bibinfo {pages} {461} (\bibinfo {year} {2021})}\BibitemShut {NoStop}%
\bibitem [{\citenamefont {Stone}\ \emph {et~al.}(2023)\citenamefont {Stone}, \citenamefont {Lu}, \citenamefont {Moille}, \citenamefont {Westly}, \citenamefont {Rahman},\ and\ \citenamefont {Srinivasan}}]{stone_wavelength-accurate_2023}%
  \BibitemOpen
  \bibfield  {author} {\bibinfo {author} {\bibfnamefont {J.~R.}\ \bibnamefont {Stone}}, \bibinfo {author} {\bibfnamefont {X.}~\bibnamefont {Lu}}, \bibinfo {author} {\bibfnamefont {G.}~\bibnamefont {Moille}}, \bibinfo {author} {\bibfnamefont {D.}~\bibnamefont {Westly}}, \bibinfo {author} {\bibfnamefont {T.}~\bibnamefont {Rahman}},\ and\ \bibinfo {author} {\bibfnamefont {K.}~\bibnamefont {Srinivasan}},\ }\bibfield  {title} {\bibinfo {title} {Wavelength-accurate nonlinear conversion through wavenumber selectivity in photonic crystal resonators},\ }\href {https://doi.org/10.1038/s41566-023-01326-6} {\bibfield  {journal} {\bibinfo  {journal} {Nature Photonics}\ ,\ \bibinfo {pages} {1}} (\bibinfo {year} {2023})},\ \bibinfo {note} {publisher: Nature Publishing Group}\BibitemShut {NoStop}%
\bibitem [{\citenamefont {Lu}\ \emph {et~al.}(2022)\citenamefont {Lu}, \citenamefont {Chanana}, \citenamefont {Zhou}, \citenamefont {Davanco},\ and\ \citenamefont {Srinivasan}}]{lu_kerr_2022}%
  \BibitemOpen
  \bibfield  {author} {\bibinfo {author} {\bibfnamefont {X.}~\bibnamefont {Lu}}, \bibinfo {author} {\bibfnamefont {A.}~\bibnamefont {Chanana}}, \bibinfo {author} {\bibfnamefont {F.}~\bibnamefont {Zhou}}, \bibinfo {author} {\bibfnamefont {M.}~\bibnamefont {Davanco}},\ and\ \bibinfo {author} {\bibfnamefont {K.}~\bibnamefont {Srinivasan}},\ }\bibfield  {title} {\bibinfo {title} {Kerr optical parametric oscillation in a photonic crystal microring for accessing the infrared},\ }\href {https://doi.org/10.1364/OL.462494} {\bibfield  {journal} {\bibinfo  {journal} {Optics Letters}\ }\textbf {\bibinfo {volume} {47}},\ \bibinfo {pages} {3331} (\bibinfo {year} {2022})}\BibitemShut {NoStop}%
\bibitem [{\citenamefont {Black}\ \emph {et~al.}(2022)\citenamefont {Black}, \citenamefont {Brodnik}, \citenamefont {Liu}, \citenamefont {Yu}, \citenamefont {Carlson}, \citenamefont {Zang}, \citenamefont {Briles},\ and\ \citenamefont {Papp}}]{black_optical-parametric_2022}%
  \BibitemOpen
  \bibfield  {author} {\bibinfo {author} {\bibfnamefont {J.~A.}\ \bibnamefont {Black}}, \bibinfo {author} {\bibfnamefont {G.~M.}\ \bibnamefont {Brodnik}}, \bibinfo {author} {\bibfnamefont {H.}~\bibnamefont {Liu}}, \bibinfo {author} {\bibfnamefont {S.-P.}\ \bibnamefont {Yu}}, \bibinfo {author} {\bibfnamefont {D.~R.}\ \bibnamefont {Carlson}}, \bibinfo {author} {\bibfnamefont {J.}~\bibnamefont {Zang}}, \bibinfo {author} {\bibfnamefont {T.~C.}\ \bibnamefont {Briles}},\ and\ \bibinfo {author} {\bibfnamefont {S.~B.}\ \bibnamefont {Papp}},\ }\bibfield  {title} {\bibinfo {title} {Optical-parametric oscillation in photonic-crystal ring resonators},\ }\href {https://doi.org/10.1364/OPTICA.469210} {\bibfield  {journal} {\bibinfo  {journal} {Optica}\ }\textbf {\bibinfo {volume} {9}},\ \bibinfo {pages} {1183} (\bibinfo {year} {2022})}\BibitemShut {NoStop}%
\bibitem [{\citenamefont {Moille}\ \emph {et~al.}(2023)\citenamefont {Moille}, \citenamefont {Lu}, \citenamefont {Stone}, \citenamefont {Westly},\ and\ \citenamefont {Srinivasan}}]{moille_fourier_2023}%
  \BibitemOpen
  \bibfield  {author} {\bibinfo {author} {\bibfnamefont {G.}~\bibnamefont {Moille}}, \bibinfo {author} {\bibfnamefont {X.}~\bibnamefont {Lu}}, \bibinfo {author} {\bibfnamefont {J.}~\bibnamefont {Stone}}, \bibinfo {author} {\bibfnamefont {D.}~\bibnamefont {Westly}},\ and\ \bibinfo {author} {\bibfnamefont {K.}~\bibnamefont {Srinivasan}},\ }\bibfield  {title} {\bibinfo {title} {Fourier synthesis dispersion engineering of photonic crystal microrings for broadband frequency combs},\ }\href {https://doi.org/10.1038/s42005-023-01253-6} {\bibfield  {journal} {\bibinfo  {journal} {Communications Physics}\ }\textbf {\bibinfo {volume} {6}},\ \bibinfo {pages} {1} (\bibinfo {year} {2023})},\ \bibinfo {note} {publisher: Nature Publishing Group}\BibitemShut {NoStop}%
\bibitem [{\citenamefont {Liu}\ \emph {et~al.}(2024)\citenamefont {Liu}, \citenamefont {Brodnik}, \citenamefont {Zang}, \citenamefont {Carlson}, \citenamefont {Black},\ and\ \citenamefont {Papp}}]{liu2024threshold}%
  \BibitemOpen
  \bibfield  {author} {\bibinfo {author} {\bibfnamefont {H.}~\bibnamefont {Liu}}, \bibinfo {author} {\bibfnamefont {G.~M.}\ \bibnamefont {Brodnik}}, \bibinfo {author} {\bibfnamefont {J.}~\bibnamefont {Zang}}, \bibinfo {author} {\bibfnamefont {D.~R.}\ \bibnamefont {Carlson}}, \bibinfo {author} {\bibfnamefont {J.~A.}\ \bibnamefont {Black}},\ and\ \bibinfo {author} {\bibfnamefont {S.~B.}\ \bibnamefont {Papp}},\ }\bibfield  {title} {\bibinfo {title} {Threshold and laser conversion in nanostructured-resonator parametric oscillators},\ }\href@noop {} {\bibfield  {journal} {\bibinfo  {journal} {Physical Review Letters}\ }\textbf {\bibinfo {volume} {132}},\ \bibinfo {pages} {023801} (\bibinfo {year} {2024})}\BibitemShut {NoStop}%
\bibitem [{\citenamefont {Brodnik}\ \emph {et~al.}(2025)\citenamefont {Brodnik}, \citenamefont {Liu}, \citenamefont {Carlson}, \citenamefont {Black},\ and\ \citenamefont {Papp}}]{brodnik_nanopatterned_2025}%
  \BibitemOpen
  \bibfield  {author} {\bibinfo {author} {\bibfnamefont {G.~M.}\ \bibnamefont {Brodnik}}, \bibinfo {author} {\bibfnamefont {H.}~\bibnamefont {Liu}}, \bibinfo {author} {\bibfnamefont {D.~R.}\ \bibnamefont {Carlson}}, \bibinfo {author} {\bibfnamefont {J.~A.}\ \bibnamefont {Black}},\ and\ \bibinfo {author} {\bibfnamefont {S.~B.}\ \bibnamefont {Papp}},\ }\bibfield  {title} {\bibinfo {title} {Nanopatterned parametric oscillators for laser conversion beyond an octave},\ }\href {https://doi.org/10.1364/OPTICA.545158} {\bibfield  {journal} {\bibinfo  {journal} {Optica}\ }\textbf {\bibinfo {volume} {12}},\ \bibinfo {pages} {337} (\bibinfo {year} {2025})},\ \bibinfo {note} {publisher: Optica Publishing Group}\BibitemShut {NoStop}%
\bibitem [{\citenamefont {Zang}\ \emph {et~al.}(2024)\citenamefont {Zang}, \citenamefont {Yu}, \citenamefont {Liu}, \citenamefont {Jin}, \citenamefont {Briles}, \citenamefont {Carlson},\ and\ \citenamefont {Papp}}]{zang2024laserpower}%
  \BibitemOpen
  \bibfield  {author} {\bibinfo {author} {\bibfnamefont {J.}~\bibnamefont {Zang}}, \bibinfo {author} {\bibfnamefont {S.-P.}\ \bibnamefont {Yu}}, \bibinfo {author} {\bibfnamefont {H.}~\bibnamefont {Liu}}, \bibinfo {author} {\bibfnamefont {Y.}~\bibnamefont {Jin}}, \bibinfo {author} {\bibfnamefont {T.~C.}\ \bibnamefont {Briles}}, \bibinfo {author} {\bibfnamefont {D.~R.}\ \bibnamefont {Carlson}},\ and\ \bibinfo {author} {\bibfnamefont {S.~B.}\ \bibnamefont {Papp}},\ }\href@noop {} {\bibinfo {title} {Laser-power consumption of soliton formation in a bidirectional kerr resonator}} (\bibinfo {year} {2024}),\ \Eprint {https://arxiv.org/abs/2401.16740} {arXiv:2401.16740} \BibitemShut {NoStop}%
\bibitem [{\citenamefont {Spektor}\ \emph {et~al.}(2024)\citenamefont {Spektor}, \citenamefont {Zang}, \citenamefont {Dan}, \citenamefont {Briles}, \citenamefont {Brodnik}, \citenamefont {Liu}, \citenamefont {Black}, \citenamefont {Carlson},\ and\ \citenamefont {Papp}}]{10.1063/5.0191602}%
  \BibitemOpen
  \bibfield  {author} {\bibinfo {author} {\bibfnamefont {G.}~\bibnamefont {Spektor}}, \bibinfo {author} {\bibfnamefont {J.}~\bibnamefont {Zang}}, \bibinfo {author} {\bibfnamefont {A.}~\bibnamefont {Dan}}, \bibinfo {author} {\bibfnamefont {T.~C.}\ \bibnamefont {Briles}}, \bibinfo {author} {\bibfnamefont {G.~M.}\ \bibnamefont {Brodnik}}, \bibinfo {author} {\bibfnamefont {H.}~\bibnamefont {Liu}}, \bibinfo {author} {\bibfnamefont {J.~A.}\ \bibnamefont {Black}}, \bibinfo {author} {\bibfnamefont {D.~R.}\ \bibnamefont {Carlson}},\ and\ \bibinfo {author} {\bibfnamefont {S.~B.}\ \bibnamefont {Papp}},\ }\bibfield  {title} {\bibinfo {title} {{Photonic bandgap microcombs at 1064 nm}},\ }\href {https://doi.org/10.1063/5.0191602} {\bibfield  {journal} {\bibinfo  {journal} {APL Photonics}\ }\textbf {\bibinfo {volume} {9}},\ \bibinfo {pages} {021303} (\bibinfo {year} {2024})}\BibitemShut {NoStop}%
\bibitem [{\citenamefont {Jin}\ \emph {et~al.}(2024)\citenamefont {Jin}, \citenamefont {Lucas}, \citenamefont {Zang}, \citenamefont {Briles}, \citenamefont {Dickson}, \citenamefont {Carlson},\ and\ \citenamefont {Papp}}]{jin_bandgap-detuned_2024}%
  \BibitemOpen
  \bibfield  {author} {\bibinfo {author} {\bibfnamefont {Y.}~\bibnamefont {Jin}}, \bibinfo {author} {\bibfnamefont {E.}~\bibnamefont {Lucas}}, \bibinfo {author} {\bibfnamefont {J.}~\bibnamefont {Zang}}, \bibinfo {author} {\bibfnamefont {T.}~\bibnamefont {Briles}}, \bibinfo {author} {\bibfnamefont {I.}~\bibnamefont {Dickson}}, \bibinfo {author} {\bibfnamefont {D.}~\bibnamefont {Carlson}},\ and\ \bibinfo {author} {\bibfnamefont {S.~B.}\ \bibnamefont {Papp}},\ }\href {https://doi.org/10.48550/arXiv.2404.11733} {\bibinfo {title} {The bandgap-detuned excitation regime in photonic-crystal resonators}} (\bibinfo {year} {2024}),\ \bibinfo {note} {arXiv:2404.11733 [physics]}\BibitemShut {NoStop}%
\bibitem [{\citenamefont {Sayson}\ \emph {et~al.}(2019)\citenamefont {Sayson}, \citenamefont {Bi}, \citenamefont {Ng}, \citenamefont {Pham}, \citenamefont {Trainor}, \citenamefont {Schwefel}, \citenamefont {Coen}, \citenamefont {Erkintalo},\ and\ \citenamefont {Murdoch}}]{sayson_octave-spanning_2019}%
  \BibitemOpen
  \bibfield  {author} {\bibinfo {author} {\bibfnamefont {N.~L.~B.}\ \bibnamefont {Sayson}}, \bibinfo {author} {\bibfnamefont {T.}~\bibnamefont {Bi}}, \bibinfo {author} {\bibfnamefont {V.}~\bibnamefont {Ng}}, \bibinfo {author} {\bibfnamefont {H.}~\bibnamefont {Pham}}, \bibinfo {author} {\bibfnamefont {L.~S.}\ \bibnamefont {Trainor}}, \bibinfo {author} {\bibfnamefont {H.~G.~L.}\ \bibnamefont {Schwefel}}, \bibinfo {author} {\bibfnamefont {S.}~\bibnamefont {Coen}}, \bibinfo {author} {\bibfnamefont {M.}~\bibnamefont {Erkintalo}},\ and\ \bibinfo {author} {\bibfnamefont {S.~G.}\ \bibnamefont {Murdoch}},\ }\bibfield  {title} {\bibinfo {title} {Octave-spanning tunable parametric oscillation in crystalline kerr microresonators},\ }\href {https://doi.org/10.1038/s41566-019-0485-4} {\bibfield  {journal} {\bibinfo  {journal} {Nature Photonics}\ }\textbf {\bibinfo {volume} {13}},\ \bibinfo {pages} {701} (\bibinfo {year} {2019})}\BibitemShut {NoStop}%
\bibitem [{\citenamefont {Chembo}\ and\ \citenamefont {Menyuk}(2013)}]{chembo_spatiotemporal_2013}%
  \BibitemOpen
  \bibfield  {author} {\bibinfo {author} {\bibfnamefont {Y.~K.}\ \bibnamefont {Chembo}}\ and\ \bibinfo {author} {\bibfnamefont {C.~R.}\ \bibnamefont {Menyuk}},\ }\bibfield  {title} {\bibinfo {title} {Spatiotemporal {Lugiato}-{Lefever} formalism for {Kerr}-comb generation in whispering-gallery-mode resonators},\ }\href {https://doi.org/10.1103/PhysRevA.87.053852} {\bibfield  {journal} {\bibinfo  {journal} {Physical Review A}\ }\textbf {\bibinfo {volume} {87}},\ \bibinfo {pages} {053852} (\bibinfo {year} {2013})},\ \bibinfo {note} {publisher: American Physical Society}\BibitemShut {NoStop}%
\bibitem [{\citenamefont {Black}\ \emph {et~al.}(2021)\citenamefont {Black}, \citenamefont {Streater}, \citenamefont {Lamee}, \citenamefont {Carlson}, \citenamefont {Yu},\ and\ \citenamefont {Papp}}]{black_group-velocity-dispersion_2021}%
  \BibitemOpen
  \bibfield  {author} {\bibinfo {author} {\bibfnamefont {J.~A.}\ \bibnamefont {Black}}, \bibinfo {author} {\bibfnamefont {R.}~\bibnamefont {Streater}}, \bibinfo {author} {\bibfnamefont {K.~F.}\ \bibnamefont {Lamee}}, \bibinfo {author} {\bibfnamefont {D.~R.}\ \bibnamefont {Carlson}}, \bibinfo {author} {\bibfnamefont {S.-P.}\ \bibnamefont {Yu}},\ and\ \bibinfo {author} {\bibfnamefont {S.~B.}\ \bibnamefont {Papp}},\ }\bibfield  {title} {\bibinfo {title} {Group-velocity-dispersion engineering of tantala integrated photonics},\ }\href {https://doi.org/10.1364/OL.414095} {\bibfield  {journal} {\bibinfo  {journal} {Optics Letters}\ }\textbf {\bibinfo {volume} {46}},\ \bibinfo {pages} {817} (\bibinfo {year} {2021})}\BibitemShut {NoStop}%
\bibitem [{\citenamefont {Jung}\ \emph {et~al.}(2021)\citenamefont {Jung}, \citenamefont {Yu}, \citenamefont {Carlson}, \citenamefont {Drake}, \citenamefont {Briles},\ and\ \citenamefont {Papp}}]{jung_tantala_2021}%
  \BibitemOpen
  \bibfield  {author} {\bibinfo {author} {\bibfnamefont {H.}~\bibnamefont {Jung}}, \bibinfo {author} {\bibfnamefont {S.-P.}\ \bibnamefont {Yu}}, \bibinfo {author} {\bibfnamefont {D.~R.}\ \bibnamefont {Carlson}}, \bibinfo {author} {\bibfnamefont {T.~E.}\ \bibnamefont {Drake}}, \bibinfo {author} {\bibfnamefont {T.~C.}\ \bibnamefont {Briles}},\ and\ \bibinfo {author} {\bibfnamefont {S.~B.}\ \bibnamefont {Papp}},\ }\bibfield  {title} {\bibinfo {title} {Tantala kerr nonlinear integrated photonics},\ }\href@noop {} {\bibfield  {journal} {\bibinfo  {journal} {Optica}\ }\textbf {\bibinfo {volume} {8}},\ \bibinfo {pages} {811} (\bibinfo {year} {2021})}\BibitemShut {NoStop}%
\bibitem [{\citenamefont {Fujii}\ and\ \citenamefont {Tanabe}(2020)}]{fujii2020dispersion}%
  \BibitemOpen
  \bibfield  {author} {\bibinfo {author} {\bibfnamefont {S.}~\bibnamefont {Fujii}}\ and\ \bibinfo {author} {\bibfnamefont {T.}~\bibnamefont {Tanabe}},\ }\bibfield  {title} {\bibinfo {title} {Dispersion engineering and measurement of whispering gallery mode microresonator for kerr frequency comb generation},\ }\href@noop {} {\bibfield  {journal} {\bibinfo  {journal} {Nanophotonics}\ }\textbf {\bibinfo {volume} {9}},\ \bibinfo {pages} {1087} (\bibinfo {year} {2020})}\BibitemShut {NoStop}%
\bibitem [{\citenamefont {Black}\ \emph {et~al.}(2023)\citenamefont {Black}, \citenamefont {Newman}, \citenamefont {Yu}, \citenamefont {Carlson},\ and\ \citenamefont {Papp}}]{black2023nonlinear}%
  \BibitemOpen
  \bibfield  {author} {\bibinfo {author} {\bibfnamefont {J.~A.}\ \bibnamefont {Black}}, \bibinfo {author} {\bibfnamefont {Z.~L.}\ \bibnamefont {Newman}}, \bibinfo {author} {\bibfnamefont {S.-P.}\ \bibnamefont {Yu}}, \bibinfo {author} {\bibfnamefont {D.~R.}\ \bibnamefont {Carlson}},\ and\ \bibinfo {author} {\bibfnamefont {S.~B.}\ \bibnamefont {Papp}},\ }\bibfield  {title} {\bibinfo {title} {Nonlinear networks for arbitrary optical synthesis},\ }\href@noop {} {\bibfield  {journal} {\bibinfo  {journal} {Physical Review X}\ }\textbf {\bibinfo {volume} {13}},\ \bibinfo {pages} {021027} (\bibinfo {year} {2023})}\BibitemShut {NoStop}%
\bibitem [{\citenamefont {Kippenberg}\ \emph {et~al.}(2004)\citenamefont {Kippenberg}, \citenamefont {Spillane},\ and\ \citenamefont {Vahala}}]{kippenberg_kerr-nonlinearity_2004}%
  \BibitemOpen
  \bibfield  {author} {\bibinfo {author} {\bibfnamefont {T.~J.}\ \bibnamefont {Kippenberg}}, \bibinfo {author} {\bibfnamefont {S.~M.}\ \bibnamefont {Spillane}},\ and\ \bibinfo {author} {\bibfnamefont {K.~J.}\ \bibnamefont {Vahala}},\ }\bibfield  {title} {\bibinfo {title} {Kerr-{Nonlinearity} {Optical} {Parametric} {Oscillation} in an {Ultrahigh}- {Q} {Toroid} {Microcavity}},\ }\href {https://doi.org/10.1103/PhysRevLett.93.083904} {\bibfield  {journal} {\bibinfo  {journal} {Physical Review Letters}\ }\textbf {\bibinfo {volume} {93}},\ \bibinfo {pages} {083904} (\bibinfo {year} {2004})}\BibitemShut {NoStop}%
\bibitem [{\citenamefont {Perez}\ \emph {et~al.}(2023)\citenamefont {Perez}, \citenamefont {Moille}, \citenamefont {Lu}, \citenamefont {Stone}, \citenamefont {Zhou},\ and\ \citenamefont {Srinivasan}}]{perez_high_performance_2023}%
  \BibitemOpen
  \bibfield  {author} {\bibinfo {author} {\bibfnamefont {E.~F.}\ \bibnamefont {Perez}}, \bibinfo {author} {\bibfnamefont {G.}~\bibnamefont {Moille}}, \bibinfo {author} {\bibfnamefont {X.}~\bibnamefont {Lu}}, \bibinfo {author} {\bibfnamefont {J.}~\bibnamefont {Stone}}, \bibinfo {author} {\bibfnamefont {F.}~\bibnamefont {Zhou}},\ and\ \bibinfo {author} {\bibfnamefont {K.}~\bibnamefont {Srinivasan}},\ }\bibfield  {title} {\bibinfo {title} {High-performance {Kerr} microresonator optical parametric oscillator on a silicon chip},\ }\href {https://doi.org/10.1038/s41467-022-35746-9} {\bibfield  {journal} {\bibinfo  {journal} {Nature Communications}\ }\textbf {\bibinfo {volume} {14}},\ \bibinfo {pages} {1} (\bibinfo {year} {2023})}\BibitemShut {NoStop}%
\bibitem [{\citenamefont {Briles}\ \emph {et~al.}(2020)\citenamefont {Briles}, \citenamefont {Yu}, \citenamefont {Drake}, \citenamefont {Stone},\ and\ \citenamefont {Papp}}]{briles_generating_2020}%
  \BibitemOpen
  \bibfield  {author} {\bibinfo {author} {\bibfnamefont {T.~C.}\ \bibnamefont {Briles}}, \bibinfo {author} {\bibfnamefont {S.-P.}\ \bibnamefont {Yu}}, \bibinfo {author} {\bibfnamefont {T.~E.}\ \bibnamefont {Drake}}, \bibinfo {author} {\bibfnamefont {J.~R.}\ \bibnamefont {Stone}},\ and\ \bibinfo {author} {\bibfnamefont {S.~B.}\ \bibnamefont {Papp}},\ }\bibfield  {title} {\bibinfo {title} {Generating {Octave}-{Bandwidth} {Soliton} {Frequency} {Combs} with {Compact} {Low}-{Power} {Semiconductor} {Lasers}},\ }\href {https://doi.org/10.1103/PhysRevApplied.14.014006} {\bibfield  {journal} {\bibinfo  {journal} {Physical Review Applied}\ }\textbf {\bibinfo {volume} {14}},\ \bibinfo {pages} {014006} (\bibinfo {year} {2020})}\BibitemShut {NoStop}%
\end{thebibliography}%

\end{document}